\documentclass[final,3p,times,twocolumn]{elsarticle}
 \biboptions{comma,sort&compress}
\usepackage{here}
 \usepackage{graphicx}
  \usepackage{epsfig}

\usepackage{amsmath}
\allowdisplaybreaks[2]

\def\nin{\noindent}
\def\beq{\begin{equation}}
\def\eeq{\end{equation}}
\def\bea{\begin{eqnarray}}
\def\eea{\end{eqnarray}}

\def\sW{{\scriptscriptstyle W}}

\journal{Nuc. Phys. (Proc. Suppl.)}

\begin{document}

\begin{frontmatter}



\title{Neutral pseudoscalar meson decays: $\pi^0 \to \gamma\gamma$ and
$\eta\to\gamma\gamma$ in $S\!U(3)$ limit$^{**}$}

 \author[label1]{Johan Bijnens}
  \address[label1]{Department of Astronomy and Theoretical Physics, Lund
University,\\S\"olvegatan 14A, SE 223-62 Lund, Sweden.}

 \author[label1,label2]{Karol Kampf\corref{cor1}}
  \address[label2]{Charles University, Faculty of Mathematics and Physics,\\
V Hole\v{s}ovi\v{c}k\'ach 2, Prague, Czech Republic}
\cortext[cor1]{Speaker}
\cortext[label3]{Supported by European Commission RTN network,
Contract MRTN-CT-2006-035482  (FLAVIAnet),
European Community-Research Infrastructure
Integrating Activity
(HadronPhysics2, Grant Agreement
n. 227431)
and the Swedish Research Council.}


\begin{abstract}
\noindent
Present and planned experiments motivate new theoretical study of properties of
light unflavoured pseudoscalar meson decays.
An overview including details on two-loop calculation in $S\!U(3)$ limit
is given.

\end{abstract}

\begin{keyword}
chiral perturbation theory \sep radiative decay of $\pi^0$ \sep higher-order
correction


\end{keyword}

\end{frontmatter}


\section{Introduction}
\nin

We would like to study unflavoured decays of light neutral pseudoscalar mesons.
This reduces the particle content to $\pi^0$, $\eta$ and eventually $\eta'$,
ruling out $K^0$ decays that violate hypercharge conservation and are
suppressed by $G_F^2$ (two-photon decays are further suppressed by $\alpha^2$
compared to hadronic ones). Standard model is thus reduced to QCD (extended
eventually only by QED corrections) which is successfully described by an
effective theory known as chiral perturbation theory (ChPT).

The $\pi^0$ meson being the lightest meson cannot decay to other hadronic
states. Its dominant decay mode (with more than 98\% probability) is $\pi^0
\to \gamma\gamma$ and is connected with the Adler-Bell-Jackiw triangle
anomaly~\cite{ABJ}. The $\pi^0\gamma\gamma$ vertex is closely connected with
other allowed $\pi^0$ decay modes: $e^+e^-\gamma$, $e^+e^-e^+e^-$, $e^+e^-$
(with branching ratios~\cite{pdg}: $0.01174(35)$, $3.34(16)\times 10^{-5}$,
$6.46(33)\times 10^{-8}$, respectively). In order to describe these processes
with sufficient precision one can employ two-flavour ChPT at appropriate order.
This can simply incorporate corrections to the current algebra result attributed
either to $m_{u,d}$ masses or electromagnetic corrections with other effects
hidden in the low energy constants (LECs). Naively, two-flavour ChPT should
converge very fast and next-to-leading order (NLO) should be sufficient from the
point of view of today's experiments. 
However, as we are exploring the anomalous sector which is poorly known,
phenomenologically richer $S\!U(3)$ ChPT must be also used in order
to obtain numerical prediction for low energy constants. This on the other hand
enables to describe $\eta\to\gamma\gamma$ in the same framework. 

The motivation for our study is both theoretical and experimental. 
As mentioned, $\pi^0\to\gamma\gamma$ represents (probably) the most
important example of the triangle anomaly in quantum field theory. It is
interesting that at NLO the amplitude gets no chiral corrections from the
so-called chiral logarithms \cite{donbij}
and this motivate the calculation at NNLO even for $S\!U(2)$ ChPT as was done in
\cite{km}.   
It was found that there are indeed chiral logarithms generated by
two-loop diagrams, but they are relatively small. It turns one's attention back
to
NLO order and contributions proportional to LECs. To this end the phenomenology
of $\eta\to\gamma\gamma$ and inevitably $\eta-\eta'$ mixing must be
employed. We intend to do the full two-loop calculation of both $\pi^0\to
\gamma\gamma$ and
$\eta\to\gamma\gamma$ in three-flavour ChPT. As a first step we will present here
the calculation and result in the $S\!U(3)$ limit, i.e. for
$m_u=m_d=m_s$.

From the experimental side let us mention the PrimEx experiment at
JLab. It is designed to perform the most precise measurement of the neutral pion
lifetime using the Primakoff effect (for first run results see
\cite{primex}). After JLab's 12 GeV upgrade the extension of the experiment
for $\eta$ and $\eta'$ radiative width measurements is planned. Neutral pion
decay modes were studied with interesting results at KTeV and it is promising to
measure them in forthcoming NA62 at CERN.

\section{Chiral expansion}
\nin
Let us briefly summarize main points of ChPT, for details see \cite{gl}.
Starting point is the chiral symmetry of QCD, called chiral because
it acts differently on left and right-handed quarks,
which is exact for $m_{u,d,s}=0$:
$$
G = S\!U(3)_L \times S\!U(3)_R,
$$
where we dropped $U(1)_A$ which is not a good symmetry due the anomaly.
However, this anomaly is proportional to a divergence which must thus vanish in
any order of perturbation theory. We are touching the problem referred as $U(1)$
problem and we will avoid further discussion assuming that the ninth axial
current is really not conserved and a possible divergence term is not present in
QCD Lagrangian (referred itself as strong CP problem). Assuming further
confinement it can be proven that the axial subgroup of $G$ is spontaneously
broken
and the associated 8 Goldstone bosons can be identified with pions, kaons and
eta.
The real non-zero masses of $u,d,s$ quarks, explicit symmetry
breaking, are added as a perturbation and this expansion around the chiral limit
together with the momentum expansion is referred to as ChPT. Standard power
counting assumes that $m_{u,d,s} = O(p^2)$, and Lorentz invariance implies
that only even powers of derivatives ($p$) can occur. The leading order (LO)
thus starts at $O(p^2)$ and one can have only tree diagrams. The next-to-leading
order (NLO) is $O(p^4)$ and can include one-loop contribution and similarly
next-to-next-to-leading order (NNLO) is $O(p^6)$ and can have up to two-loop
diagrams. The last important point to be discussed here is the so-called chiral
or external anomaly which would correctly incorporate the full symmetry pattern
of QCD. It is connected with the fact that quarks carry also electromagnetic
charge. In fact some Green functions of QCD (e.g. $VVA$) are not invariant under
chiral symmetry, the difference was calculated first by Bardeen \cite{ABJ}
and incorporated to the action by Wess, Zumino and Witten (WZW) \cite{WZW}.
This action starts at $O(p^4)$ and thus the anomalous vertex shifts our
counting by one order (i.e. NNLO here is $O(p^8)$).

                      
\section{Decay modes}
\nin
We are primarily interested now in two-photon decays of $\pi^0$ and
$\eta$. Nevertheless let us summarize shortly their ``spin-off'' products,
namely
\begin{itemize}
\item $\pi^0 \to e^+e^-\gamma$ so called Dalitz decay is important in
  normalization of rare pion and kaon decays. This was supported by its
  precise and stable prediction: for 30 years its official PDG value was same
(based on LAMPF experiment). However the last edition changed this number, based
on ALEPH results and so it will have impact in other measurements via the
  normalization. The differential decay rate is discussed in~\cite{kkn}.
\item $\pi^0 \to e^+e^-e^+e^-$ or double Dalitz decay enables experimental
verification of $\pi^0$
  parity. KTeV set recently new limits on parity and CPT violation
\cite{ktev08}
\item $\pi^0\to e^+e^-$  depends directly only on fully off-shell
  $\pi^0\gamma^*\gamma^*$ vertex. KTeV measurement \cite{ktev06} is off by
$3.5\sigma$ from
  the existing models. It can set
  valuable limits on models beyond SM
\item $\pi^0 \to invisible(\gamma)$, exotics and violation processes were also
studied in $\pi^0$ decays. It includes mainly decay to neutrinos but is also
interesting in beyond SM scenarios (neutralinos, extra-light neutral vector
particle, etc.) 
\end{itemize}
(for more references cf.~\cite{pdg}).
The same modes are also possible in $\eta$ decays, see e.g.~\cite{persson}.

\section{LO and NLO calculation}
\begin{figure}[hbt]
\begin{center}
\epsfig{file=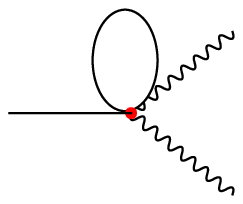}\qquad\epsfig{file=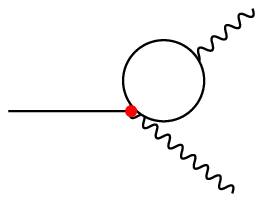}
\end{center}
\caption{\scriptsize One-loop corrections to two photon pseudoscalar decays.
A (red) dot represents the odd-parity coupling.}
\label{oneloop}
\end{figure}

\nin
In the chiral limit the decay width is fixed by axial anomaly with the result
\begin{equation}
 \Gamma(\pi^0 \to \gamma\gamma)^{CA} =
\frac{m_{\pi^0}^3}{64\pi}\biggl(\frac{\alpha N_C}{3\pi F_\pi} \biggr)^2 \approx
7.76\,\rm{eV}.
\end{equation}
It is in excellent agreement with experiment, which is the opposite situation to
two-photon $\eta$ decay. In $S\!U(3)$ limit (and also in chiral limit) the
two studied amplitudes are connected by Wigner-Eckart theorem $\sqrt3 T_{\eta}
= T_{\pi^0}$, i.e.
\begin{equation}
 \Gamma(\eta \to \gamma\gamma)^{CA} =
\frac{m_{\eta}^3}{64\pi}\biggl(\frac{\alpha
N_C}{3\sqrt3\pi F_\pi} \biggr)^2 \approx 173\,{\rm eV},
\end{equation}
which is far from experiment $0.510\pm0.026\,{\rm keV}$ \cite{pdg}. (Note that
using
$F_\eta$ instead of $F_\pi$ makes this difference even larger.) The difference
is attributed to $\eta-\eta'$ mixing. At NLO order, apart from tree diagrams
coming from WZW and $O(p^6)$ odd-parity Lagrangian, we should include two
one-loop
topologies (depicted in Fig.\ref{oneloop}).

The full one-loop calculation based on wave function renormalization 
and chiral expansion of masses and decay constants leads to:
\begin{align}
 &\Gamma(\pi^0 \to \gamma\gamma)^{NLO} = \Gamma(\pi^0 \to \gamma\gamma)^{CA}
\times \Bigr[1 - \frac{256\pi^2}{3}  m_\pi^2 C^{Wr}_7\Bigl]^2,\notag\\
&\Gamma(\eta \to \gamma\gamma)^{NLO} = \Gamma(\eta \to \gamma\gamma)^{CA}
\times \Bigr[\frac{F_\pi^2}{F_\eta^2} + \frac{256\pi^2}{9}\label{Gnlo}\\ 
&\qquad\times\Bigl((4 m_K^2 - 7 m_\pi^2) C^{Wr}_7 + 24 (m_K^2 - m_\pi^2)
C^{Wr}_8 \Bigr)
\Bigl]^2.\notag
\end{align}
Note, as anticipated, the very simple, polynomial form of the results without
logarithms. This is especially accomplished by correct replacement of $F_0$,
i.e. $F_0 \to$ $F_\pi$ and $F_\eta$ in $\pi^0$ and $\eta$ decay respectively.

It is clear from~(\ref{Gnlo}) that $\eta-\eta'$ mixing must be hidden in
$C^W_8$ LEC. A rough estimate using resonance saturation suggests that
$C^W_8$ must be much bigger than $C^W_7$. For further discussion see \cite{am}
and \cite{km}.


\section{Two-loop calculation in $SU(3)$ limit}
\nin
The $O(p^8)$, (or equivalently NNLO, or two-loop) calculation was already
performed for $\pi^0\to\gamma\gamma$ in two-flavour ChPT. Natural extension
for $S\!U(3)$ will supply us with both $\pi^0$ and also $\eta\to\gamma\gamma$
and enable to test and verify chiral expansion in odd intrinsic
sector (cf. study for even sector~\cite{hansilaria}). It is, however, clear
that this calculation will be difficult: we are facing instead of one, three
different scales
in overlapping two-loop diagrams (sunset and vertex). Big effort was already
given in the simpler two-point (sunset) case, and we still lack general
analytic form. We plan to calculate it using method described
in~\cite{kl4} but we need to go beyond the loop integrals computed there. 
There exists, however, apart from chiral limit, one non-trivial limit which can
be
used to obtain analytical result as it depends again only on one scale. It is an
$S\!U(3)$ limit, where we set $m_u = m_d = m_s=m \neq 0$. This we
can simply connect with $O(p^2)$ mass: $M_\pi^{(0)2} = 2 B m$.    

The current algebra prediction, fixed by the anomaly, is free from any mass
contribution. The mass enters explicitly at NLO order only, and therefore to
obtain NNLO order we need to connect $O(p^2)$ parameter with
physical (referring to a world where $m_u=m_d=m_s$) $S\!U(3)$
mass:
$$
\frac{M_\pi^2}{M^2} = 1+ \frac{M_\pi^2}{F_\pi^2}\Bigl[ \frac{L}{3} - 8 (3 L_4^r
+ L_5^r - 6 L_6^r- 2 L_8^r) \Bigr] + O(M_\pi^4)
$$
with chiral logarithm defined as $(4\pi)^2L=\ln{M_\pi^2/\mu^2}$. On the other
hand connection of $F_0$ with physical $S\!U(3)$ decay constants is
needed up to NNLO order
$$
\frac{F_\pi}{F_0} = 1 + \frac{M_\pi^2}{F_\pi^2}( 12 L_4^r + 4 L_5^r - \frac32 L)
+ \frac{M_\pi^4}{F_\pi^4} f_{NNLO} + O(M_\pi^6).
$$
The NNLO part was already calculated in general $S\!U(N_F)$ in \cite{lujiehans}
and for our $N_F=3$ is given by
$$
f_{NNLO}= \frac{\lambda_F}{(4\pi)^2} + \bar\lambda_F + {\cal K}_F + r_F
+\frac{1561 L}{288(4\pi)^2}- \frac{421}{2304(4\pi)^4}
$$
with
\begin{align*}
 &\lambda_F = -2 L_1^r - 9 L_2^r - 7/3 L_3^r\\
&\bar\lambda_F = 8 (3 L_4^r+L_5^r) (21 L_4^r + 7 L_5^r- 24 L_6^r - 8 L_8^r)\\
& {\cal K}_F = 1/2 (34 {\cal K}_1 + 13 {\cal K}_2 + 13 {\cal K}_3 - 45 {\cal
K}_4
- 15 {\cal K}_5)\\
& r_F = 8 ( C_{14}^r + 3 C_{15}^r + 3 C_{16}^r + C_{17}^r)
\end{align*}
and ${\cal K}_i = (4 L_i^r - \Gamma_i L) L$ using renormalization coefficients
taken from \cite{gl}.

As already mentioned, for $S\!U(3)$ limit $\pi^0$ and $\eta$ decays
are related by Wigner-Eckart theorem and we thus need to calculate only one
of these processes. Following Weinberg power-counting at NNLO we need to
consider $a$) tree graphs with either $a_1$) one vertex from odd $O(p^8)$ sector
or
$a_2$) one from odd (even) $O(p^6)$ and second from even (odd) $O(p^4)$; $b$)
one-loop diagrams with one vertex with NLO coupling (even or odd) and $c$) the
two-loop graphs with one vertex taken from the WZW Lagrangian. All other
vertices should be generated by the $O(p^2)$ chiral Lagrangian. 

\begin{figure}[hbt]
\begin{center}
\begin{center}
\epsfig{file=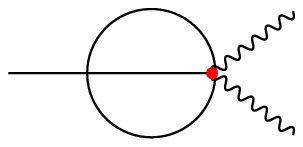}\quad\epsfig{file=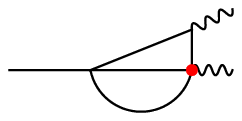}\quad
\epsfig{file=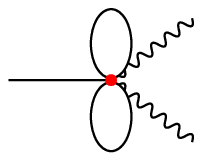}\quad\epsfig{file=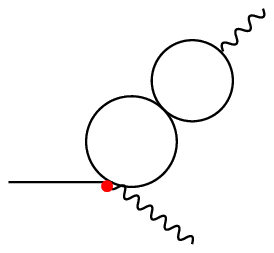}\quad\epsfig{file=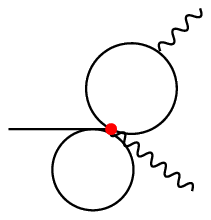}
\quad\epsfig{file=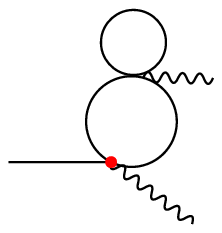}\quad\epsfig{file=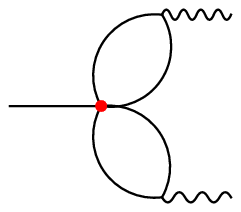}
\end{center}
\end{center}
\caption{\scriptsize Two-loop corrections to two photon pseudoscalar decays.}
\label{twoloop}
\end{figure}

Case $a_2$) is treated via wave function renormalization.
However,
the odd-sector Lagrangian at $O(p^8)$ for three flavours has not yet been
studied.
The connected LEC will be denoted as $D^W_i$ and set only a posteriori to cancel
all local divergences. Concerning one-loop Feynman diagrams, we have already
summarized them in Fig.\ref{oneloop}, for NNLO the topology stays the same, we
need
just to insert higher-order vertices. Non-trivial part of calculation is hidden
in two loops. The Feynman diagrams to deal with are summarized in
Fig.\ref{twoloop}. Corrections (tadpoles) to propagators are not depicted. Note
that the most of diagrams are the same as in the two-flavour case. As
anticipated by
the nature of the anomaly there is one new topology (the last one diagram in
Fig.\ref{twoloop}) with anomalous vertex without direct photon insertion
(so-called Chesire-cat smile). Of course, into these graphs one should insert
all possible combinations of pions, kaons and eta (fortunately in $S\!U(3)$
limit
with identical masses).

We summarize the preliminary result in the following form ($T$ is normalized as
$T^{CA}=1$
at LO, cf. eqs~(\ref{Gnlo})).
\begin{align}
\frac{F_\pi^4}{m_\pi^4} T^{NNLO} &=  \frac{\lambda}{(4\pi)^2} + (4
\pi)^2\bar\lambda + (4 \pi)^2 {\cal K} + r + \frac{329 L}{96(4\pi)^2}\notag\\
& +\frac{9\sqrt3 {\rm Cl}_2(\pi/3) -4 \zeta(3) - \frac{7093}{1152}}{(4\pi)^4}
\end{align}
with
\begin{align*}
&\lambda = 0\\
&\bar\lambda =-\frac{256}{3} F_0^2 C_7^{Wr}(3 L_4^r + L_5^r - 3 L_6^r - L_8^r)
\\
& {\cal K} = 4 {\cal K}^\sW_4 + 10 {\cal K}^\sW_7 -2 {\cal K}^\sW_9 + 4
{\cal K}^\sW_{11}-\tfrac12 {\cal K}^\sW_{13}-2 {\cal K}^\sW_{14} - {\cal
K}^\sW_{15} \\
& r = -32 C_{12}^r - 96 C_{13}^r - 4 D^{Wr}_{lim}
\end{align*}
and ${\cal K}^\sW_i = (4 F_0^2 C_i^{Wr} - \eta^{(3)}_i L) L$ using
renormalization
coefficients
taken from \cite{BGTodd}. The $O(p^8)$ chiral coupling which would cancel local
divergences in $S\!U(3)$ limit is denoted by $D^W_{lim}$ and our exact
calculation
fixes its decomposition
\begin{align*}
 D^W_{lim} = \frac{(c \mu)^{2(d-4)}}{F_0^2} \biggl[ 
D^{Wr}_{lim}(\mu) + \Lambda^2 \frac{127}{12} + \Lambda \Bigl( \frac{208}{3}
L_1^r 
+32 L_2^r\\ +\frac{248}{9} L_3^r +36 L_4^r + 12 L_5^r + \frac{91}{128(4\pi)^2}
+(4\pi F_0)^2 (8 C_4^{Wr} \\ + \frac{100}{9}C_7^{Wr} - 4 C_9^{Wr} +8 C_{11}^{Wr}
- C_{13}^{Wr} - 4 C_{14}^{Wr} -2 C_{15}^{Wr}
) \Bigr)
\biggr].
\end{align*}

\section{Conclusion}
\nin
We have summarized here our preliminary results concerning a two-loop
calculation of $\pi^0 (\eta)\to \gamma\gamma$ in $S\!U(3)$ limit (where
$m_{u,d}=m_s=m$). The word preliminary refers also to the fact that independent
calculation with physical masses is in progress~\cite{bk} and it should allow us
to crosscheck here presented result in this limit. The possibility of studying
two-photon decays of light-meson on lattice was very recently demonstrated in
\cite{lattice}. The simple analytical result can be very useful in this
direction as one can vary masses without changing LECs.

\section*{Acknowledgements}
\nin
K.K. would like to thank the organizers for a very enjoyable conference.












\end{document}